\documentstyle[preprint,aps]{revtex}
\tightenlines

\input epsf.tex
\def\DESepsf(#1 width #2){\epsfxsize=#2 \epsfbox{#1}}
\begin{document}
\preprint{\vbox{\hbox{COLO-HEP-409}\hbox{hep-ph/9807231}}}
\draft
\title{$\alpha_s$ and Gauge Coupling Unification in Flipped SU(5) Models \\
  with Gauge-Mediated Supersymmetry Breaking} 
\author{K.T. Mahanthappa$^*$ and Sechul Oh$^{\dagger}$ }
\address{Department of Physics, University of Colorado, Boulder, Colorado 80309, USA}
\date{July, 1998}
\maketitle 
\begin{abstract}
We examine gauge coupling unification and prediction for $\alpha_s (M_Z)$ in minimal flipped SU(5) with gauge-mediated supersymmetry breaking.  
We include threshold corrections at weak scale, messenger scale and unification scale, and explicitly show that in this model there can be either \emph{one-step} or 
\emph{two-step} gauge coupling unification, depending on unification-scale threshold corrections.  
The experimental value of $\alpha_s (M_Z)$ constrains heavy particle masses in one-step unification.  
In the case of two-step unification, we examine the prediction for $\alpha_s (M_Z)$ with two-loop, light, messenger and heavy threshold corrections, and find it to be compatible with the updated experimental data.
\end{abstract}
\vspace{1 cm} 
PACS numbers: 11.30.Pb, 12.10.Dm, 12.10.Kt, 12.60.Jv \\
Keywords: flipped SU(5), gauge-mediated supersymmetry breaking, gauge coupling unification, $\alpha_s$, threshold corrections  \\ \\
$*$ E-mail address: ktm@verb.colorado.edu  \\ 
$\dagger$ E-mail address: ohs@colorado.edu 

\newpage
It is a well-known fact that supersymmetry (SUSY) provides an elegant solution to the hierarchy problem in particle physics.  Further support for SUSY as an important ingredient in physics beyond the standard model (SM) is provided by the fact that the particle content of the minimal supersymmetric SM (MSSM) leads to the unification of the three gauge coupling constants of SM.  This has lead to a search for a realistic SUSY grand unified theory (GUT) \cite{1}.  
Unfortunately the simplest SUSY GUT based on SU(5) \cite{2} is ruled out as the unification of gauge coupling constants with the inclusion of threshold effects is incompatible with experimental limits on proton decay in gauge-mediated SUSY breaking (GMSB) models \cite{3,300}; in gravity-mediated models \emph{very small} parameter space is allowed \cite{4}.  The root of this problem lies in proton decay due to Higgsino exchange which leads to dimension-5 operators \cite{5}.   
The so-called supersymmetric flipped SU(5) (SUSY FS) models \cite{6,7}, based on SU(5)$\times$U(1), overcome this problem elegantly by having a simple doublet-triplet splitting mechanism which is a consequence of non-adjoint Higgses breaking SU(5)$\times$U(1) down to SM.  There exist different versions of FS models $-$ a minimal model \cite{7} and several non-minimal string-derived versions \cite{8}.  
A SUSY FS model is strictly not a grand unified model; one can have \emph{one-step} or \emph{two-step} unification.  
The one-step unification is realized by requiring the U(1) gauge coupling, $\tilde \alpha_1$, of SU(5)$\times$U(1) to be equal to the SU(5) coupling $\alpha_5$ at a scale $M_{32}$ (at which SU(3) and SU(2) couplings of SM, $\alpha_2$ and $\alpha_3$, are equal to $\alpha_5$), whereas in two-step unification $\tilde \alpha_1$ and $\alpha_5$ get unified at a scale greater than $M_{32}$.  This entails a relationship between the two couplings which could have its origin in string theory. 
Extensive analyses of FS models have been performed using gravity-mediated SUSY breaking \cite{7,8}.  Our purpose in this paper is to study the minimal FS model using GMSB with emphasis on prediction for $\alpha_s(M_Z)$ and its sensitivity to the representation content of the messenger sector.  

The minimal SUSY FS model \cite{7} has the following particle content:  this model has three generations of quarks and leptons 
\begin{eqnarray}
F_i ({\bf 10}, 1) &=& \{Q, d^c, \nu^c \}, \;\; 
\bar f_i (\bar {\bf 5}, -3) = \{L, u^c \}, \\ \nonumber
l_i^c ({\bf 1}, 5) &=& e^c,  \;\;\;\;\;  (i=1,2,3) 
\end{eqnarray}
a conjugate pair of ten-dimensional Higgses to break SU(5)$\times$U(1) down to 
SU(3)$_C \times$SU(2)$_L \times$U(1)$_Y$ 
\begin{eqnarray}
H ({\bf 10}, 1) = \{Q_H, d_H^c, \nu_H^c \}, \;\; 
\bar H (\overline{{\bf 10}}, -1) = \{Q_{\bar H}, d_{\bar H}^c, \nu_{\bar H}^c \},
\end{eqnarray}
and a conjugate pair of five-dimensional Higgses to break the electroweak symmetry SU(2)$_L\times$U(1)$_Y$ 
\begin{eqnarray}
h ({\bf 5}, -2) = \{ H_2, H_3 \}, \;\;  \bar h (\bar {\bf 5}, 2) = \{ \bar H_2, \bar H_3 \},
\end{eqnarray}
and in addition some singlet fields $\phi_m ({\bf 1},0)$.
Here the numbers in parentheses denote the dimensions of the SU(5) representations and the U(1) quantum number of the corresponding matter fields under SU(5)$\times$U(1) and the contents in curly brackets represent states in the corresponding matter multiplets.  The states in the $H$ multiplet are labeled by the same symbols as in the $F_i$ multiplet, with an additional subscript $H$.  The states $H_2$ and $H_3$ in the $h$ multiplet are the doublet Higgs and the color-triplet Higgs, respectively.  
The superpotential of the minimal SUSY FS is given by 
\begin{eqnarray}
W &=& \lambda_1^{ij} F_i F_j h + \lambda_2^{ij} F_i \bar f_j \bar h 
  +\lambda_3^{ij} \bar f_i l_j^c h +\lambda_4 HHh 
  +\lambda_5 \bar H \bar H \bar h  \\ \nonumber 
 &+&\lambda_6^{im} F_i \bar H \phi_m +\lambda_7^m h \bar h \phi_m 
  +\lambda_8^{mnp} \phi_m \phi_n \phi_p.  
\end{eqnarray}
Among several interesting features in this model is a natural doublet-triplet splitting mechanism.  The neutral components $\nu_H^c$ and $\nu_{\bar H}^c$ of ${\bf 10}$ and $\overline{{\bf 10}}$ Higgs multiplets $H$ and $\bar H$ acquire large equal vacuum expectation values (VEVs) $< \nu_H^c > =< \nu_{\bar H}^c > = v$.  From this VEVs the color-triplet Higgses $H_3$ and $\bar H_3$ obtain heavy Dirac masses $M_{H_3}= \lambda_4 v$ and $M_{\bar H_3}= \lambda_5 v$ by coupling to $d_H^c$ and $d_{\bar H}^c$ in $H$ and $\bar H$ multiplets via the superpotential couplings $\lambda_4 HHh$ and $\lambda_5 \bar H \bar H \bar h$, while leaving the doublet Higgses light.  Also the $X$, $Y$ gauge bosons and gauginos in adjoint representation and the $Q_H$ and $Q_{\bar H}$ Higgs bosons and Higgsinos in ${\bf 10}$ and $\overline{{\bf 10}}$ representations acquire equal masses $M_V = g_5 v$.    
 
In GMSB models \cite{9} messenger fields transmit SUSY breaking to the fields of visible sector via loop diagrams involving SU(3)$_C \times$SU(2)$_L \times$U(1)$_Y$ gauge interactions.  The simplest model consists of messenger fields which transform as a single flavor of vectorlike ${\bf 5} +\bar {\bf 5}$ of SU(5).  These messenger fields may be coupled to a SM singlet chiral superfield $S$ through the superpotential 
\begin{eqnarray}
W_{messenger} = \lambda_D S D \bar D + \lambda_L S L \bar L, 
\end{eqnarray} 
where the fields have the SM representations and quantum numbers 
$D:({\bf 3},{\bf 1})_{Y=-2/3}$, 
$\bar D:(\bar {\bf 3},{\bf 1})_{Y=2/3}$, $L:({\bf 1},{\bf 2})_{Y=-1}$, and $\bar L:({\bf 1},{\bf 2})_{Y=1}$.  
The scalar and $F$ components of $S$ acquire VEVs $<S>$ and $<F_S>$, respectively, through their interactions with the fields of hidden sector, which results in breakdown of SUSY.  
It is known that for messenger fields in complete SU(5) representation, at most 
four (${\bf 5} +\bar {\bf 5}$) pairs, or one (${\bf 5} +\bar {\bf 5}$) and one (${\bf 10} +\overline{{\bf 10}}$) pair are allowed to ensure that the gauge couplings remain perturbative up to the GUT scale \cite{10}. 

In our analysis we will take the messenger fields to transform in complete SU(5) representations, which means that the messenger fields have the SU(5) representations and the U(1) quantum numbers 
$({\bf 5},0)$, $(\bar {\bf 5},0)$, $({\bf 10},0)$, and $(\overline{{\bf 10}},0)$, respectively, under the FS gauge group SU(5)$\times$U(1). 
The radiatively generated soft SUSY-breaking masses of gaugino and scalars, $\tilde M_i$ and $\tilde m^2$, at messenger scale $M$ are given by \cite{11,1100} 
\begin{eqnarray}
\tilde{M_i} (M) &=& (n_5 +3 n_{10}) g \left( {\Lambda \over M} \right) 
  {\alpha_i(M) \over 4\pi} \Lambda, 
\label{Mi}
\end{eqnarray}
\begin{eqnarray}
\tilde{m}^2 (M) &=& 2 (n_5 +3 n_{10}) f \left( {\Lambda \over M} \right)   
  \sum_{i=1}^3 k_i C_i \left( {\alpha_i(M) \over 4\pi} \right)^2 \Lambda^2, 
\label{m2}
\end{eqnarray}
where the parameter $\Lambda$ is defined by $\Lambda = <F_S> / <S>$.  
The messenger scale $M$ is given by $M = \lambda <S>$ with $\lambda$ a universal Yukawa coupling in the messenger sector at GUT scale.  
$\alpha_i$ $(i=1,2,3)$ are the three SM gauge couplings with GUT normalization for $\alpha_1$.  The $k_i$ are 1, 1, 3/5 for SU(3), SU(2) and U(1)$_Y$, respectively.  The $C_i$ are zero for gauge singlets, and 4/3, 3/4 and $(Y/2)^2$ for the fundamental representations of SU(3), SU(2) and U(1)$_Y$, respectively 
(with $Y$ defined by $Q= I_3 + Y/2$).  $n_5$ and $n_{10}$ denote the number of (${\bf 5} +\bar {\bf 5}$) and (${\bf 10} +\overline{{\bf 10}}$) pairs, respectively.  
$g(x)$ and $f(x)$ are messenger scale threshold functions with $x= \Lambda / M$. 
We require that electroweak symmetry be radiately broken.  
We use the input values $\alpha_s(M_Z)=0.118$, $\sin^2 \theta_W (M_Z)=0.2315$ and $\alpha(M_Z)=1/128$. Using the appropriate renormalization group equations (RGEs) \cite{12}, we first go up to the messenger scale $M$ with gauge and Yukawa couplings, and fix the sparticle masses with the boundary conditions (\ref{Mi}) and (\ref{m2}).  We next go down with the $6 \times 6$ mass matrices for the squarks and sleptons to find the sparticle spectrum. 
We take $\Lambda$ to be around 100 TeV to ensure that the sparticle masses are of the order of the weak scale. The upper bound on the gravitino mass of about $10^4$ eV restricts $1 < M / \Lambda < 10^4$ \cite{1100,13}.  

In SUSY FS, gauge coupling unification is not automatic, but it can be achieved in a certain range of masses of heavy particles at the GUT scale if one includes all threshold corrections at the weak scale, the messenger scale and the GUT scale.  First we investigate this \emph{one-step} gauge coupling unification in the minimal SUSY FS with GMSB.  This analysis gives upper bounds on masses of the heavy particles like the color-triplet Higgs bosons, depending on input values of $\alpha_s (M_Z)$.  
By running the gauge couplings up to the GUT scale, we determine the GUT-scale mismatch $\delta \alpha_3^{-1} \equiv \alpha_3^{-1} (M_{GUT}) - \alpha_5^{-1} (M_{GUT})$, where $\alpha_5^{-1} (M_{GUT})$ is defined by $\alpha_5^{-1} (M_{GUT}) \equiv \alpha_1^{-1} (M_{GUT}) =\alpha_2^{-1} (M_{GUT})$.  
We find that the GUT-scale mismatch is given by \cite{3} 
\begin{eqnarray}
\delta \alpha_3^{-1} = \delta_{weak} +\delta_{messenger} +\delta_{GUT}, 
\label{alpha3} 
\end{eqnarray}
where 
\begin{eqnarray}
\delta_{weak} = -{1 \over 4 \pi} \left( 4 \ln {m_{\tilde g} \over m_{\tilde w}}
  -{8 \over 5} \ln {m_{\tilde h} \over m_t} \right), 
\label{weak}
\end{eqnarray}
\begin{eqnarray} 
\delta_{messenger} = -{1 \over 4 \pi} \left( {12 \over 5} n_5 \ln {M_D \over M_L}
  -{18 \over 5} n_{10} \ln {M_Q \over M_U} -{6 \over 5} n_{10} \ln {M_Q \over 
  M_E} \right), 
\label{messenger}
\end{eqnarray} 
\begin{eqnarray}
\delta_{GUT} = -{1 \over 4 \pi} \left( {12 \over 5} \ln {M_{H_3} M_{\bar H_3}
  \over M_{GUT}^2} \right). 
\label{GUT}
\end{eqnarray} 
The term $\delta_{weak}$ is the weak-scale threshold correction which depends on masses of gluino $\tilde g$, wino $\tilde w$ and higgsino $\tilde h$.  Here we have omitted the negligible contribution from scalar particles.  The term  $\delta_{messenger}$ is the messenger-scale threshold correction which depends on the mass splitting of messenger fields of $n_5$  
$({\bf 5} + \bar {\bf 5})$ and $n_{10}$ $({\bf 10} + \overline{{\bf 10}})$ pairs.  The messenger fields $D$ and $L$ in $({\bf 5} + \bar {\bf 5})$ representation have the representations and quantum numbers $({\bf 3},{\bf 1})_{Y=-2/3}$ and 
$({\bf 1},{\bf 2})_{Y=-1}$, respectively, under the SM gauge group.  Similarly the fields $Q$, $U$ and $E$ in $({\bf 10} + \overline{{\bf 10}})$ representation have the representations and quantum numbers $({\bf 3},{\bf 2})_{Y=1/3}$, 
$({\bf 3},{\bf 1})_{Y=4/3}$ and $({\bf 1},{\bf 1})_{Y=-2}$, respectively.  The term  $\delta_{GUT}$ is the GUT-scale threshold correction which depends on masses of the color-triplet Higgs bosons $H_3$ and $\bar H_3$ in the Higgs pentaplets.  Here we have assumed that $M_V \approx M_{GUT}$.   
We note that in comparison with the case of minimal SUSY SU(5), in minimal SUSY FS $\delta_{GUT}$ has the different form as in Eq.(\ref{GUT}) because of the different heavy particle content, while $\delta_{weak}$ and $\delta_{messenger}$ have the same form as those of minimal SUSY SU(5) \cite{3}.  
It has been shown \cite{3,300} that in the minimal SUSY SU(5) with GMSB, the GUT-scale threshold correction is not consistent with the constraints from proton decay and $b-\tau$ Yukawa coupling unification.  
In particular, the limit from proton decay implies the color-triplet Higgs mass $M_{H_3} \geq M_{GUT}$ \cite{14}.  
However, in minimal SUSY FS, there is no such constraint on $M_{H_3}$ and 
$M_{\bar H_3}$, because of absence of stringent constraint from proton decay and  absence of $b-\tau$ Yukawa coupling unification in this model unless the Yukawa couplings $\lambda_{4,5}$ are extremely small.  
(It is interesting to get lower limit on the magnitude of $\lambda_4$ and $\lambda_5$ from the present limits on proton decay.  In FS the dimension-five Higgsino exchange term giving rise to proton decay is proportional to $\lambda^2 m_{H_3 \bar H_3} / M^2_{H_3} (\mbox{or}\; M^2_{\bar H_3})$, where $\lambda$ is the Yukawa coupling(s) giving rise to  fermion mass(es) and can at most be of $O(1)$, $m_{H_3 \bar H_3} = \lambda_7 <\phi> \sim m_W$, and $M^2_{H_3} = (\lambda_4 / g_5) M_{GUT}$, $M^2_{\bar H_3} = (\lambda_5 / g_5) M_{GUT}$ \cite{7}.  This, combined with present limits on life time of proton decay, implies that $\lambda_4$ and $\lambda_5$ cannot be less than $O(10^{-7})$.)
Thus it is possible for $\delta_{GUT}$ to cancel out both $\delta_{weak}$ and $\delta_{messenger}$ in a certain region of $M_{H_3}$ and $M_{\bar H_3}$ such as 
$M_{H_3, \bar H_3} < M_{GUT}$, which results in \emph{one-step} unification.     
 
Since one can estimate all masses in Eqs.(\ref{weak}), (\ref{messenger}), (\ref{GUT}) except for the color-triplet Higgs masses and determine $\delta \alpha_3^{-1}$ by running the two-loop RGEs, one can find the upper bounds on $M_{H_3} M_{\bar H_3}$ from Eq.(\ref{alpha3}).  This approach is similar to one in the case of the minimal SUSY SU(5) presented in Ref.\cite{3}.  We use $m_t=$ 175 GeV and $m_{\tilde g}/ m_{\tilde w} = \alpha_3 / \alpha_2 \approx 3.5$.  To find the upper bound on $M_{H_3} M_{\bar H_3}$, we take $m_{\tilde h} \approx $ 1 TeV.  The mass ratio of the messenger fields $M_D /M_L$, $M_Q /M_U$ and $M_Q /M_E$ can be calculated by running the relevant Yukawa couplings in RGEs.  
The numerical values used in our analysis are shown in the paragraphs below Eq.(\ref{delheavy}). 
In Fig. 1, we show upper bounds on $M_{H_3}$ versus $\alpha_s (M_Z)$ for the cases without and with messenger fields, assuming $M_{H_3} =M_{\bar H_3}$.  
The solid line is the case of the minimal SUSY FS with \emph{no} messenger fields and the dashed line is the case with two $({\bf 5} + \bar {\bf 5})$ pairs in the messenger sector.
The dotted-dashed line is the case with one $({\bf 10} + \overline{{\bf 10}})$ pair.  
The color-triplet Higgs mass $M_{H_3}$ is less than $M_{GUT}$ for the whole range of values of $\alpha_s (M_Z)$ shown in Fig.1.  

In the case of \emph{two-step} unification, the prediction for the strong coupling $\alpha_s (M_Z)$ was studied based on gravity-mediated SUSY breaking in Ref.\cite{15}.  We now investigate the effect of gauge-mediated SUSY breaking on the prediction for $\alpha_s (M_Z)$ in minimal SUSY FS.  The main difference in our analysis arises from the effect of the messenger sector in GMSB, as well as the mass spectrum of sparticles in GMSB.  

The prediction for $\alpha_s (M_Z)$ in minimal SUSY FS is given by \cite{15} 
\begin{eqnarray}
\alpha_s (M_Z) = {(7/3) \alpha \over 5 ( \sin^2 \theta_W -\tilde \delta ) -1 
  + (11 / 2\pi) \alpha \ln(M_{32}^{max} / M_{32})}, 
\label{alphas}
\end{eqnarray} 
where $\alpha$ is the electromagnetic coupling and $\theta_W$ is the weak mixing angle.  
Here $\tilde \delta$ denote the corrections to $\sin^2 \theta_W$, which is given by 
\begin{eqnarray}
\tilde \delta = \tilde \delta_{2loop} +\tilde \delta_{light} +\tilde \delta_{messenger} +\tilde \delta_{heavy}. 
\end{eqnarray}
The first term $\tilde \delta_{2loop}$ is two-loop correction having 
$\tilde \delta_{2loop} \approx 0.0030$.  The terms $\tilde \delta_{light}$ and 
$\tilde \delta_{messenger}$ are threshold corrections at the weak and messenger scale, while $\tilde \delta_{heavy}$ is heavy particle threshold correction similar to the GUT-scale threshold correction in the minimal SUSY SU(5). 
$M_{32}$ denotes a first unification scale in \emph{two-step} unification scenario of this model, at which the SU(3) and SU(2) gauge couplings become equal.  At this scale the SM U(1)$_Y$ gauge coupling $\alpha_1$ with GUT normalization evolves in general to a different value $\alpha_1^{\prime}$.  
Above the scale $M_{32}$, the governing gauge group is SU(5)$\times$U(1) whose U(1) gauge coupling $\tilde \alpha_1$ is related to $\alpha_1^{\prime}$ and the SU(5) gauge coupling $\alpha_5$ by 
\begin{eqnarray}
25 \alpha_1^{\prime -1} = \alpha_5^{-1} + 24 \tilde \alpha_1^{-1}. 
\end{eqnarray}
Above the scale $M_{32}$ the couplings $\alpha_5$ and $\tilde \alpha_1$ evolve to finally become equal at a higher scale $M_{51}$.  
The maximum possible value of $M_{32}$, $M_{32}^{max}$, is given by 
\begin{eqnarray}
\alpha_1^{-1} -\alpha_5^{-1} = {b_1 \over 2\pi} 
  \ln {M_{32}^{max} \over M_Z}, 
\end{eqnarray}
where $b_1$ is a beta function coefficient (with GUT normalization) for the SM U(1)$_Y$. 

We now present a detailed calculation of the threshold corrections.  
The threshold corrections can be obtained from the general formula \cite{16} 
\begin{eqnarray}
\tilde \delta_i = {\alpha \over 20\pi} \sum_{R_j} C(R_j) \ln {M_j \over M_i},  
\end{eqnarray}
where 
\begin{eqnarray}
C(R) = {10 \over 3} b_1(R) -8 b_2(R) + {14 \over 3} b_3(R).  
\end{eqnarray}
Here $i$ denotes the corresponding scale and the sum runs over all the corresponding scale particles $R_j$ with masses $M_j$.  $C(R)$ is a linear combination of the one-loop beta function coefficients of the representation $R$.  We find that the threshold corrections are given by \cite{15}
\begin{eqnarray}
\tilde \delta_{light} &=& {\alpha \over 20\pi} \left( -3 \ln {m_t \over M_Z} +{28 \over 3} 
   \ln  {m_{\tilde g} \over M_Z} -{32 \over 3} \ln {m_{\tilde w} \over M_Z} 
   - \ln {m_H \over M_Z} -4 \ln {m_{\tilde h} \over M_Z} \right. \\ \nonumber
  &+& \left. {5 \over 2} \ln {m_{\tilde q} \over M_Z} 
   -3 \ln {m_{\tilde l_L} \over M_Z}
   +2 \ln {m_{\tilde l_R} \over M_Z} 
   -{35 \over 36} \ln {m_{\tilde t_1} \over M_Z} 
   -{19 \over 36} \ln {m_{\tilde t_2} \over M_Z} \right),  
\end{eqnarray}
\begin{eqnarray}
\tilde \delta_{messenger} = {\alpha \over 20\pi} \left( 6 n_5 \ln {M_D \over M_L}
  -10 n_{10} \ln {M_Q \over M_U} -4 n_{10} \ln {M_Q \over M_E} \right), 
\label{delmessenger} 
\end{eqnarray}
\begin{eqnarray}
\tilde \delta_{heavy} = {\alpha \over 20\pi} \left( -6 \ln {M_{32} \over M_{H_3}}
   -6 \ln {M_{32} \over M_{H_3}} +4 \ln {M_{32} \over M_V} \right). 
\label{delheavy}
\end{eqnarray}

To estimate $\tilde \delta_{light}$, we calculate sparticle mass spectrum using the GMSB model.  We obtain the mass spectrums for each case of one 
$({\bf 5} +\bar {\bf 5})$ pair only (i.e., $n_5=1$ and $n_{10}=0$) and  one 
$({\bf 10} +\bar {\bf 10})$ pair only (i.e., $n_5=0$ and $n_{10}=1$) as follows.  In the case of $n_5=1 (0)$ and 
$n_{10}=0 (1)$, gluino mass $m_{\tilde g}= 557 (667)$ GeV, wino mass  
$m_{\tilde w}= 167 (200)$ GeV, heavy Higgs mass $m_{H}= 313 (273)$ GeV, 
higgsino mass $m_{\tilde h}= 307 (266)$ GeV, stop masses $m_{\tilde t_1}= 727 (611)$ GeV, $m_{\tilde t_2}= 654 (524)$ GeV, squark masses $m_{\tilde q}= 700 (567)$ GeV, and slepton masses $m_{\tilde l_L}= 222 (176)$ GeV, $m_{\tilde l_R}= 112 (91)$ GeV.  Here we have used the parameters $\Lambda = 60 (25)$ TeV, 
$M = 2 (20) \Lambda$, $\tan \beta = 20 (20)$, $m_t = 175 (175)$ GeV.  
Also, we have used the parameter $\mu < 0$, where $\mu$ is the Higgs mixing parameter in the superpotential. 
In Fig. 2 and 3 we show the prediction for $\alpha_s (M_Z)$ in the minimal FS with GMSB as a function of $M_{32} / M_{32}^{max}$.  
Fig. 2 and 3 correspond the case with $n_5=1$ and $n_{10}=0$ and with $n_5=0$ and $n_{10}=1$ in the messenger sector, respectively.  
The updated experimental value $\alpha_s (M_Z) = 0.118 \pm 0.003 \; ( 1 \sigma )$ \cite{alphas}. 
The solid lines are the cases with \emph{no} threshold corrections ($\tilde \delta_{light} =\tilde \delta_{messenger} =\tilde \delta_{heavy} =0$).  We use the updated experimental value $\sin^2 \theta_W (M_Z) = 0.2315 \pm 0.0004$. 
The effect of $\tilde \delta_{light}$ is represented by the dashed lines ( i.e.,  $\tilde \delta_{light} \neq 0$, $\tilde \delta_{messenger} =
\tilde \delta_{heavy} =0$ ).
In both cases, one $({\bf 5} +\bar {\bf 5})$ pair only and one $({\bf 10} +\overline{{\bf 10}})$ pair only, the light threshold correction $\tilde \delta_{light} > 0$.  So the effect of $\tilde \delta_{light}$ tends to a little bit increase the prediction for $\alpha_s (M_Z)$.  This result is similar to that in the case of the gravity-mediated models.  

The messenger threshold correction arises from the mass splitting of the messenger fields.  
We note that in Eq.(\ref{delmessenger}) the contribution from the $({\bf 10} +\overline{{\bf 10}})$ pair $(n_{10} = 1)$ has the negative sign, while the contribution from the $({\bf 5} +\bar {\bf 5})$ pairs has the positive sign.  Thus if the mass ratios of the messenger fields in Eq.(\ref{delmessenger}) are larger than 1, the effect of the $({\bf 10} +\overline{{\bf 10}})$ pair tends to decrease the value of $\alpha_s (M_Z)$, while the effect of the $({\bf 5} +\bar {\bf 5})$ pairs tends to increase $\alpha_s (M_Z)$.  
This threshold correction from messenger fields is a new effect from the GMSB, which does not exist in the gravity-mediated models.  The contribution of $\tilde \delta_{messenger}$ is comparable with or even larger than that of $\tilde \delta_{light}$. 
For instance, in the case with one $({\bf 10} +\overline{{\bf 10}})$ pair only, the threshold effect of the messenger fields can compensate the light threshold effect $\tilde \delta_{light} > 0$ and the sum effect of $\tilde \delta_{light} +\tilde \delta_{messenger}$ can reduce the prediction for $\alpha_s (M_Z)$.   
Using the same parameter values as in the case of sparticle masses, we obtain the mass ratios of the messenger fields: $M_D / M_L =1.39$ for $n_5=1$ and $n_{10}=0$, and $M_Q / M_U =1.44$ and $M_Q / M_E =3.61$ for $n_5=0$ and $n_{10}=1$.  
In Fig. 2 and 3 the dotted-dashed lines refer to the case with both the messenger and light threshold effects ( i.e., $\tilde \delta_{light} \neq 0$, $\tilde \delta_{messenger} \neq 0$, $\tilde \delta_{heavy} =0$ ).  In the case of $n_5=0$ and $n_{10}=1$, $\tilde \delta_{messenger} < 0$, while in the case of $n_5=1$ and $n_{10}=0$, $\tilde \delta_{messenger} > 0$. 
In Fig. 3 we see that the threshold effect of one 
$({\bf 10} +\overline{{\bf 10}})$ pair lowers the prediction for $\alpha_s (M_Z)$.  

The effect of the heavy thresholds are the same as that in the case of the gravity-mediated models, since the messenger scale in the GMSB is much lower than the scale of the heavy particle masses such as $M_{H_3}$, $M_{\bar H_3}$ and $M_V$.  Thus just as in the case of the gravity-mediated models, it is possible that $M_{H_3}, \; M_{\bar H_3} < M_V \approx M_{32}$, since there is no stringent constraint on $M_{H_3}$, $M_{\bar H_3}$ from proton decay.  This case corresponds to the heavy threshold effect $\tilde \delta_{heavy} < 0$ and leads to the prediction of lower $\alpha_s (M_Z)$.  

In conclusion, we have investigated gauge coupling unification and the prediction for $\alpha_s (M_Z)$ in the minimal FS with GMSB.  
In our analysis we have included the weak-scale, the messenger-scale, and the GUT-scale threshold corrections, and have explicitly shown that in this model there can be either \emph{one-step} or \emph{two-step} gauge coupling unification, depending on the GUT-scale threshold corrections.  
The experimental value of $\alpha_s (M_Z)$ constrains heavy particle masses like the color-triplet Higgses in \emph{one-step} unification.  
In the case of \emph{two-step} unification, we have examined the prediction for $\alpha_s (M_Z)$ with the two-loop, light, messenger and heavy threshold corrections.  The effect of the light thresholds tends to increase the value of 
$\alpha_s (M_Z)$, which is similar to that in the case of the gravity-mediated models.  However, the messenger threshold effect which does not exist in the gravity-mediated models can increase or decrease the value of $\alpha_s (M_Z)$, 
and can compensate the light threshold correction, depending on the chosen representation of the messenger sector.   
Since the messenger scale in the GMSB is much lower than the scale of the heavy particle masses, the heavy threshold effect is expected to be the same as that in the gravity-mediated models and can lead to the prediction for lower $\alpha_s (M_Z)$. 
Including all the threshold corrections, we have shown that the prediction for $\alpha_s (M_Z)$ is compatible with the updated experimental data. \\

We would like to thank D.V. Nanopoulos for clarifying remarks concerning flipped SU(5) models.  We also thank B. Dutta for helpful discussions. \\
This work was supported in part by the US Department of Energy Grant No. DE FG03-95ER40894.

\newpage
\hspace*{5 cm}
FIGURE CAPTIONS \\ 
\begin{itemize}
\item[Fig. 1~:] {Upper bound on the color-triplet Higgs mass $M_{H_3}$ as a function of $\alpha_s (M_Z)$, assuming $M_{H_3}=M_{\bar H_3}$.  All the threshold corrections are included. 
The solid line corresponds to the case of the minimal SUSY FS with \emph{no} messenger fields.  The dashed line and dotted-dashed line correspond to the cases with two $(5 +\bar 5)$ pairs and one $(10 +\bar 10)$ pair of messenger fields, respectively. } 
 
\item[Fig. 2~:] {The prediction for $\alpha_s (M_Z)$ in minimal SUSY FS with GMSB as a function of the ratio $R \equiv M_{32}/ M_{32}^{max}$, in the case of 
$n_5=1$ and $n_{10}=0$ of messenger fields.  The solid lines correspond to the 
range of predictions for $\sin^2 \theta_W =0.2315 \pm 0.0004$ with \emph{no} threshold corrections.  The dashed lines refer to the case with light threshold correction only and the dotted-dashed lines refer to the case with both the messenger and light threshold corrections.  Here heavy threshold correction is not included.  Note that the case $R=1$ corresponds to the minimal SUSY SU(5) prediction. The updated experimental value $\alpha_s (M_Z) = 0.118 \pm 0.003 \; ( 1 \sigma )$.}  

\item[Fig. 3~:] {The same as Fig.2, except $n_5=0$ and $n_{10}=1$.} 
\end{itemize}

\begin{figure}[htb]
\vspace{1 cm}

\centerline{ \DESepsf(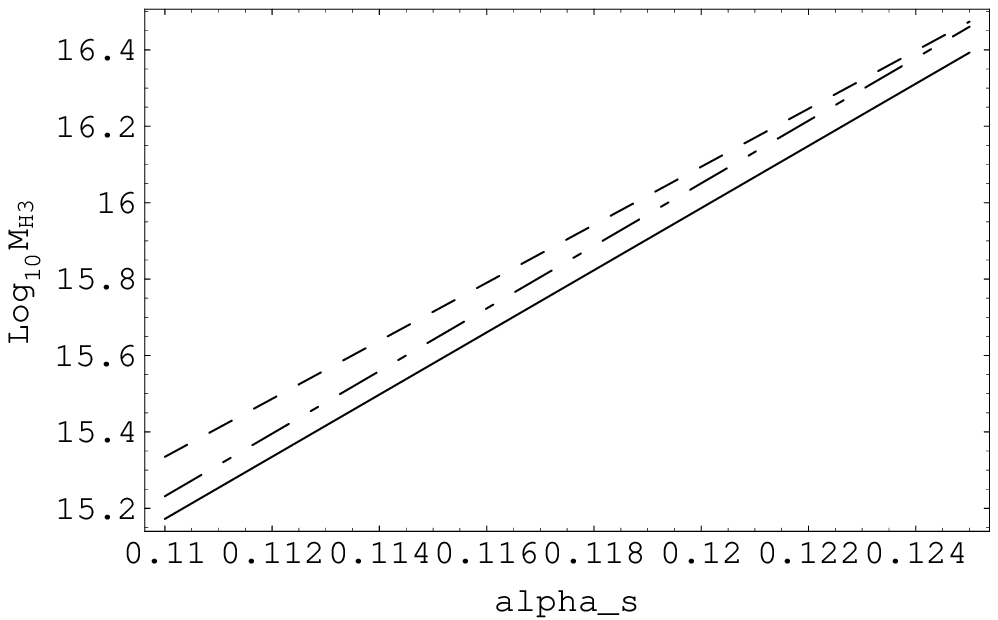 width 12 cm) }
\smallskip
\caption {}
\vspace{1 cm}

\centerline{ \DESepsf(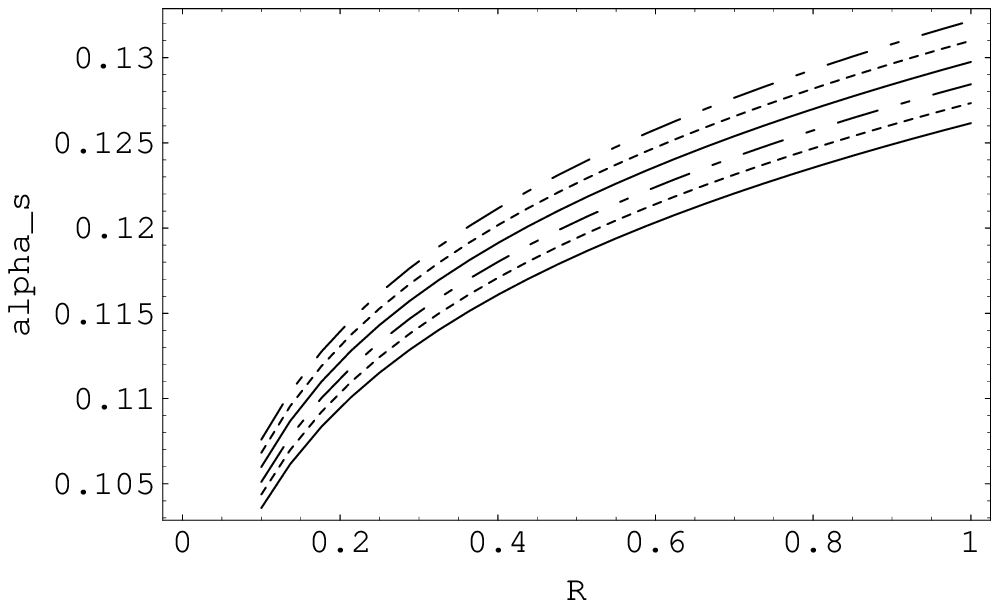 width 12 cm) }
\smallskip
\caption {}
\vspace{1 cm}

\centerline{ \DESepsf(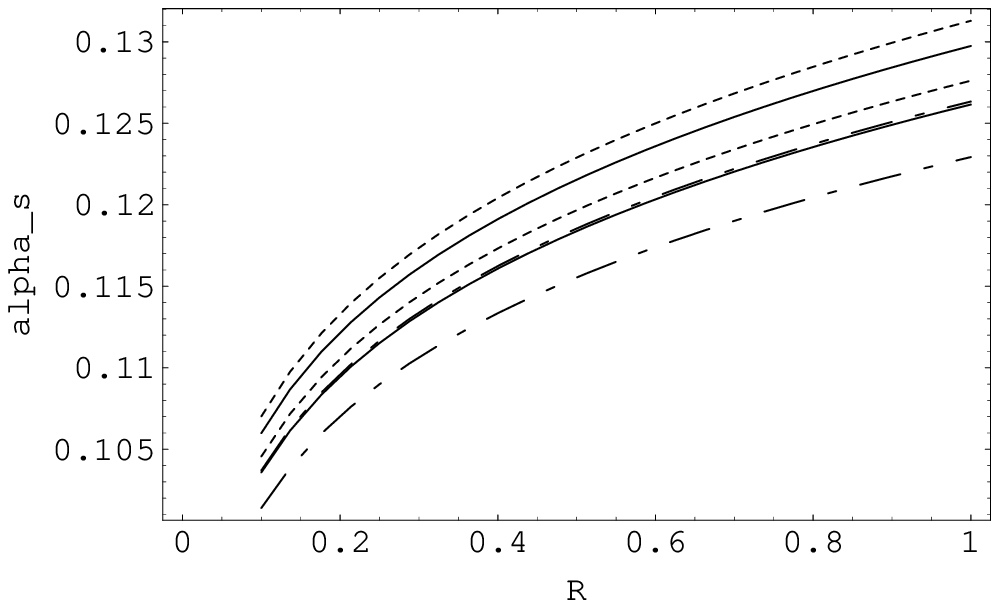 width 12 cm) }
\smallskip
\caption {}
\vspace{3 cm}
\end{figure}


\newpage
\begin{thebibliography}{[001]}

\bibitem{1} For a recent review see R.N. Mohapatra, TASI 97 Proceedings;  
 hep-ph/9801235.
\bibitem{2} S. Dimopoulos and H. Georgi, Nucl. Phys. {\bf B193}, 150 (1981);  N. Sakai, Z. Phys. {\bf C11}, 153 (1981).      
\bibitem{3} C.D. Carone and H. Murayama, Phys. Rev. {\bf D53}, 1658 (1996). 
\bibitem{300} J.A. Bagger, K.T. Matchev, D.M. Pierce, and R.-J. Zhang, Phys. Rev. Lett. {\bf 78}, 1002 (1997). 
\bibitem{4} J. Hisano, T. Moroi, K. Tobe and T. Yanagida, Mod. Phys. Lett. {\bf A10}, 2267 (1995); J. Hisano, H. Murayama and T. Yanagida, Nucl. Phys. {\bf B402}, 46 (1993).
\bibitem{5} N. Sakai and T. Yanagida, Nucl. Phys. {\bf B197}, 533 (1982); 
S. Weinberg, Phys. Rev. {\bf D53}, 1658 (1996).
\bibitem{6} S.M. Barr, Phys. Lett. {\bf B112}, 219 (1982); J.P. Derendinger, J.E. Kim and D.V. Nanopoulos, Phys. Lett. {\bf B139}, 170 (1984).
\bibitem{7} I. Antoniadis, J. Ellis, J. Hagelin and D.V. Nanopoulos, Phys. Lett. {\bf B194}, 231 (1987); J. Ellis, J. Hagelin, S. Kelly and D.V. Nanopoulos, Nucl. Phys. {\bf B311}, 1 (1983).
\bibitem{8} For a review and references see J.L. Lopez and D.V. Nanopoulos, hep-ph/9511266.  
\bibitem{9} For a review and references see G.F. Giudice and R. Rattazzi, hep-ph/9801271 (Submitted to Phys. Rep.).  
\bibitem{10} T. Moroi, H. Murayama and T. Yanagida, Phys. Rev. {\bf D48}, 2995 (1993); B. Brahmachari, U. Sarkar and K. Sridhar,  Mod. Phys. Lett. {\bf A8}, 3349 (1993).
\bibitem{11} S. Dimopoulos, S. Thomas and J.D. Wells, Nucl. Phys. {\bf B488}, 39 (1997); S.P. Martin, Phys. Rev. {\bf D55}, 3177 (1997). 
\bibitem{1100} J.A. Bagger, K.T. Matchev, D.M. Pierce, and R.-J. Zhang, Phys. Rev. {\bf D55}, 3188 (1997).   
\bibitem{12} V. Barger, M. Berger, P. Ohmann, and R.J.N. Phillips, Phys. Rev. {\bf D51}, 2438 (1995) and references therein. 
\bibitem{13} H. Pagels and J.R. Primack, Phys. Rev. Lett. {\bf 48}, 223 (1982). 
\bibitem{14} B.D. Wright, hep-ph/9404217. 
\bibitem{15} J. Ellis, J.L. Lopez and D.V. Nanopoulos, Phys. Lett. {\bf B371}, 65 (1996).
\bibitem{16} J. Ellis, S. Kelly and D.V. Nanopoulos, Nucl. Phys. {\bf B373}, 55 (1992). 
\bibitem{alphas} Particle Data Group, Phys. Rev. {\bf D54}, 65 (1996).
\end{thebibliography}
\end{document}